\documentclass[aip,apl,
amsmath,amssymb,
reprint,%
superscriptaddress,
]{revtex4-1}

\usepackage{graphicx}
\usepackage{dcolumn}
\usepackage{bm}

\usepackage[utf8]{inputenc}
\usepackage[T1]{fontenc}
\usepackage{mathptmx}
\begin{document}

\title[]{\hspace{12.0mm}$\delta$-PVDF Based Flexible Nanogenerator}

\author{Varun Gupta}
\affiliation{Quantum Materials and Devices Unit, Institute of Nano Science and Technology, Knowledge City, Sector 81, Mohali 140306, India}
\author{Anand Babu}
\affiliation{Quantum Materials and Devices Unit, Institute of Nano Science and Technology, Knowledge City, Sector 81, Mohali 140306, India}
\author{Sujoy Kumar Ghosh}
\affiliation{Department of Physics, Jadavpur University, Kolkata 700032, India}
\affiliation{Present address: NEST, Istituto Nanoscienze-CNR and Scuola Normale Superiore, Piazza San Silvestro 12, I-56127 Pisa, Italy}
\author{Zinnia Mallick}
\affiliation{Quantum Materials and Devices Unit, Institute of Nano Science and Technology, Knowledge City, Sector 81, Mohali 140306, India}
\author{Hari Krishna Mishra}
\affiliation{Quantum Materials and Devices Unit, Institute of Nano Science and Technology, Knowledge City, Sector 81, Mohali 140306, India}
\author{Dipankar Mandal}
\email{dmandal@inst.ac.in}
\affiliation{Quantum Materials and Devices Unit, Institute of Nano Science and Technology, Knowledge City, Sector 81, Mohali 140306, India}

\begin{abstract}

Delta ($\delta$) phase comprising polyvinylidene fluoride (PVDF) nanoparticles are fabricated through electrospray technique by applying 0.1 MV/m electric field at ambient temperature and pressure, which is 10$^{3}$ times lower than the typical value, required for $\delta$-phase transformation. The X-ray diffraction (XRD) and selected area electron diffraction (SAED) patterns are clearly indicating the $\delta$-phase formation. The piezo- and ferro- electric response of the $\delta$-PVDF nanoparticles has been demonstrated through scanning probe microscopic technique based on piezoresponse force microscopy (PFM). The vertical piezoelectric response, indicated by \textit{d$_{33}$} coefficient, is found $\sim$-11 pm/V. Kink propagation model is adopted to justify the $\delta$-phase conversion in electrospray system. The electrical response from $\delta$-PVDF nanoparticle comprised nanogenerator under the external impacts and acoustic signal indicates that molecular ferroelectric dipoles responsible for piezoelectric responses, are poled in-situ during nanoparticle formation, thus further electrical poling is not necessary. 
 
\end{abstract}
\maketitle

    The piezoelectric polymers have been extensively explored due to their potential applications as transducers, actuators, sensors and energy harvesters. \cite{{won2015piezoelectric},{reece2003nonvolatile},{singh2014ferroelectric},{ghosh2017bio},{martins2014electroactive},{ghosh2016high},{https://doi.org/10.1002/adma.201505684},{ranjan2007phase},{mandal2011origin},{muller2008ferroelectric}} PVDF is one of such piezoelectric semi-crystalline polymer that has been widely investigated and considered to have at least five distinct crystalline phases. \cite{lovinger1983ferroelectric} These phases have different chain conformations designated as all trans (TTTT) planar zigzag $\beta$-phase (form I), T$_{3}$GT$_{3}$G´ $\gamma$-phase (form III), $\epsilon$-phase (form V), TGTG´ (trans-gauche-trans-gauche) $\alpha$-phase (form II) and $\delta$-phase (form IV). \cite{{banik1979theory},{furukawa1989ferroelectric}} Despite of the same chain conformations of $\alpha$ and $\delta$-phase, the $\alpha$-phase is non-polar and paraelectric in nature due to its centro-symmetric \cite{{broadhurst1978piezoelectricity},{lovinger1981unit},{li2013revisiting},{hasegawa1972crystal},{hasegawa1972molecular}} (P2$_{1}$/c) unit cell. The marcomolecular chain of PVDF (chemical formula, -(-CH$_{2}$-CF$_{2}$-)$_{n}$-) consists of repeated units of -CH$_{2}$-CF$_{2}$- as a monomer. The electronegative F and electropositive H atoms are the basis of molecular dipoles, (i.e. -CH$_{2}$- and -CF$_{2}$- dipoles) perpendicular to molecular c-axis. The dipole moments inside the unit cell of $\alpha$-phase are aligned in anti-parallel direction, resulting a non-polar state of PVDF. \cite{lovinger1983ferroelectric} In contrast, $\delta$-phase exhibiting a non centro-symmetric (P2$_{1}$cn) unit cell \cite{bachmann1980crystal} that gives rise to piezoelectric, pyroelectric and ferroelectric properties. In case of $\delta$-phase, every second chain along c-axis is rotated about 180 degrees, as a result the dipole moments in unit cell of $\delta$-phase are aligned parallel to each other and perpendicular to molecular axis, giving rise to a polar state of PVDF. \cite{dvey1980dynamics} Also, the X-ray diffraction (XRD) studies have shown that the $\delta$-phase has the similar interplanar d-spacing as of $\alpha$-phase but with the significant changes in their relative intensities for the specific set of h, k, l values. In a $\delta$-phase unit cell, with an n glide perpendicular to c-axis, hk0 reflections are absent for h + k = 2n + 1. \cite{{bachmann1980crystal},{naegele1978formation}} Theoretically, the most accepted mechanism to achieve $\delta$-phase is, to rotate the each of second polymer chain about its c-axis of $\alpha$-phase unit cell.\cite{naegele1978formation} Experimentally, in 1978, the pioneer group, Davis \textit{et al.} had shown that $\delta$-phase could be achieved from $\alpha$-phase by applying a strong electrical field of $\sim$170 MV/m. \cite{davis1978electric} In subsequent years, many other groups \cite{{naegele1978formation},{lovinger1981molecular},{costa1993electric},{doi:10.1021/acsmacrolett.9b00166},{davies1979evidence}} has also demonstrated the formation of $\delta$-phase by applying higher electric field, with the coercive field of more than 100 MV/m. In addition, there have been also reported a very few different techniques such as electroforming \cite{{li2013revisiting},{davis1978electric},{li2013low}}, solid state processing \cite{martin2017solid} and quenching to achieve the $\delta$-phase \cite {{gan2016ferroelectric},{servet1979polymorphism},{qian2018unveiling},{garcia2017ferroelectric}} in PVDF (Discussion S1 and Fig. S1, supplementary material). Apart from thick and thin films of PVDF different types of structures such as rods, tubes, particles, flakes have also gained the interest due to their technological importance. \cite{fu2020interfacial}

In this work, we are reporting the formation of piezoelectric $\delta$-phase comprising PVDF (henceforth, abbreviated as $\delta$-PVDF) nanoparticles, at the lowest possible electric field reported till date (Table SI, supplementary material). The application of lower electric field gives an advantage over the dielectric fatigue and electrical breakdown of the polymer samples in comparison to higher electrical poling field. In this process, we have used the electrospray technique \cite{wilm1994electrospray} to achieve the $\delta$-PVDF nanoparticles at room temperature (Discussion S2 and Fig. S2, supplementary material).

\begin{figure}
\includegraphics[scale=0.39]{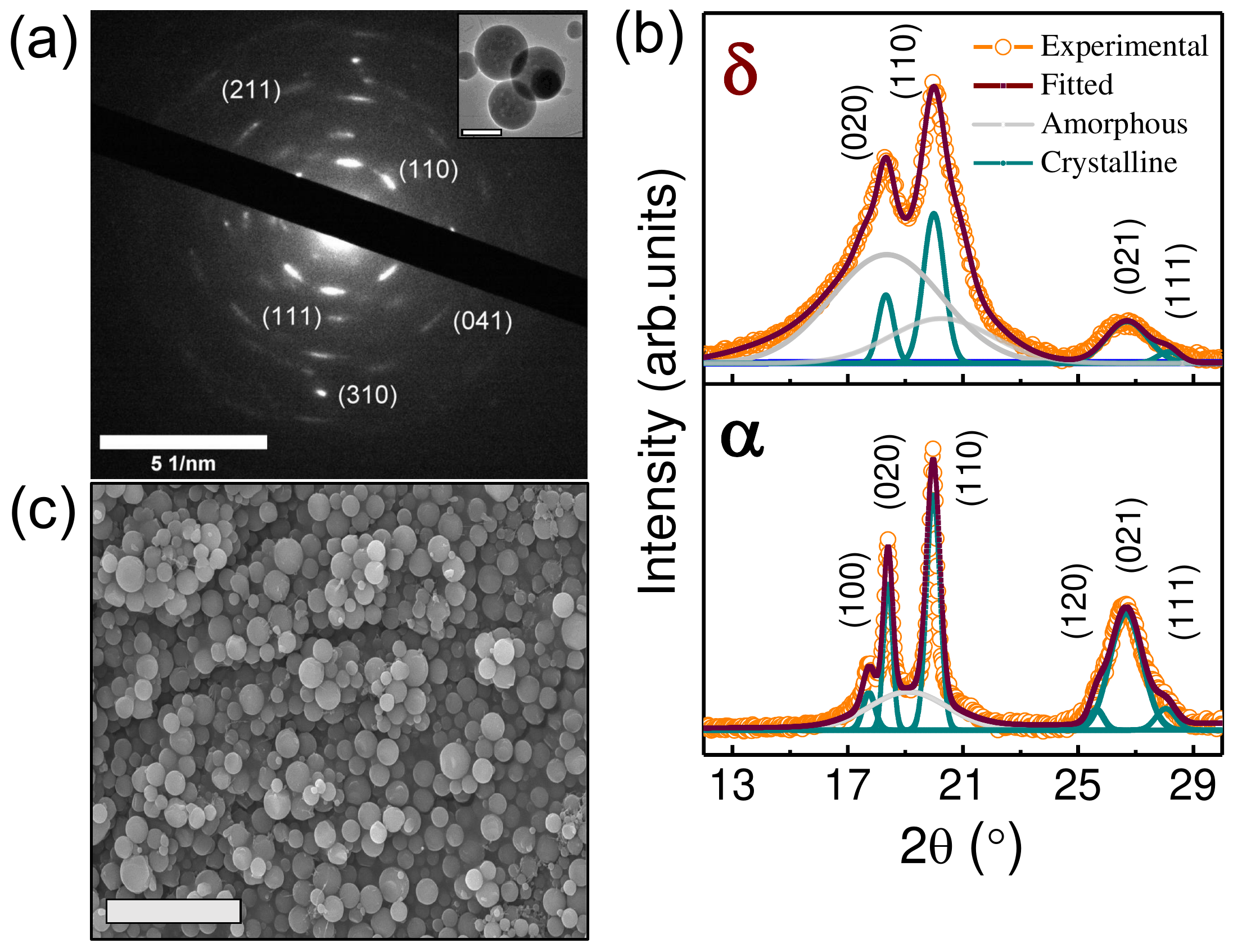}% Here is how to import EPS art
\caption{\label{fig:epsart}(a) SAED pattern of $\delta$-phase in PVDF nanoparticles. Inset shows the TEM imaging of PVDF nanoparticles (scale bar 200 nm). (b) X-ray diffraction pattern of PVDF nanoparticles in different phases. Experimental observation of $\delta$-phase at 30ºC (marked as $\delta$) and its conversion to $\alpha$-phase at 170ºC after temperature treatment (marked as $\alpha$). (c) Morphology of $\delta$-PVDF nanoparticles as uniform distribution in SEM (scale bar 5 $\mu$m).}
\end{figure}

The SAED pattern (Fig. 1(a)) obtained in transmission electron microscope (TEM) (Discussion S3, supplementary material) confirms the diffraction pattern of $\delta$-PVDF nanoparticles (inset of Fig. 1(a)) with similar interplanar spacing (d) values as obtained in XRD. The d-spacing of 0.441 nm corresponding to peak for (110) shows the brightest ring in electron diffraction. The (110) plane is consistent with the highest intensity peak in XRD. The reflections of plane (020) has not observed in the SAED, since the (110) and (020) are very closely originated and cannot be distinguished in bright rings as diffraction angles are close for these two planes. The next diffraction ring is observed for plane (111). Since (021) and (111) planes are closely spaced and have lesser intense rings due to structure factors dependency of these reflections, so only (111) was observed. The diffraction rings for plane (041), (211) and (310) was also observed in SEAD which belongs to $\delta$-phase PVDF {\cite{bachmann1980crystal} with much smaller d-spacing, for higher 2$\theta$ values above 40º. The XRD pattern of PVDF nanoparticles at room temperature (RT) in Fig. 1(b) (upper panel, marked as $\delta$), resembles precisely with the $\delta$-phase represented XRD patterns, having similar d-spacing \cite{bachmann1980crystal} and interplanar reflections of (020), (110), (021) and (111) at 2$\theta$ peak positions of 18.3º, 19.9º, 26.7º and 28.1º respectively, as earlier reported by pioneer groups \cite{{naegele1978formation},{davis1978electric}} with lattice parameters of a = 4.96 {\AA}, b = 9.64 {\AA}, c = 4.62 {\AA}. It confirms the presence of $\delta$-phase in these PVDF nanoparticles fabricated through the electrospray system. Further, we have demonstrated the reverse phase transformation of PVDF from $\delta$-phase to $\alpha$-phase   in nanoparticles. In this case, higher temperature treatment (170ºC) was given to the $\delta$-PVDF nanoparticles. The XRD pattern of the PVDF nanoparticles after heat treatment at 170ºC (Fig. 1(b) lower panel, marked as $\alpha$) shows the $\alpha$-phase PVDF characteristic diffraction peaks, appeared at (2$\theta$) 17.6º, 18.3º, 19.9º, 25.6º, 26.7º and 28.1º attribute to lattice planes of (100), (020), (110), (120), (021) and (111), respectively. \cite{{hasegawa1972molecular},{ince2010impact}} This phase change from $\delta$ to $\alpha$-phase has occurred due to the temperature treatment above the melt crystallization of PVDF, that destroys dipolar alignments of $\delta$-phase (ferroelectric) and converts it to non-polar $\alpha$-phase (paraelectric). The comparative study of both phases (Fig. 1(b)) shows that the only significant change has observed as decrease in the peak intensities of reflections (100) and (120) at 2$\theta$ position 17.6º and 25.6º in $\delta$-phase, compared to the $\alpha$-phase PVDF, with the same d-spacing and lattice parameters. So, it can be concluded that there is no disordering, contraction or change in lattice shape and size. Then the only possible changes in both the phases could be due to symmetry change of the unit cell lattice. \cite{{gan2016ferroelectric},{naegele1978formation}} This conclusion is consistent with basic unit cell structure of $\alpha$ and $\delta$-phase. That reconfirms the $\delta$-phase formation in electrospray system. The crystallite size of the $\delta$-phase and  $\alpha$-phase crystalline lamella was calculated to be $\sim$6 nm and $\sim$13 nm, respectively (Discussion S4, supplementary material). Also, the degree of crystallinity ($\chi_{c}$) of $\delta$-phase and $\alpha$-phase, from XRD pattern, was found to be 47\% and 52\%, respectively (Discussion S4, supplementary material). The fourier transform infrared spectroscopy (FTIR) study of the nanoparticles (Fig. S3 (marked as $\delta$), supplementary material) shows the IR absorption bands of $\delta$-PVDF as reported by K. Tashiro \textit{et al.}\cite{doi:10.1021/acs.macromol.0c02567}. The vibrational bands at 1182 cm$^{-1}$ and 1209 cm$^{-1}$ are one of the conclusive way to identify the $\delta$-phase through FTIR.  The IR exposure causes the symmetric ($\nu_{s}$) and anti-symmetric ($\nu_{as}$) stretching of -CF$_{2}$- dipoles at 1182 cm$^{-1}$ and 1209 cm$^{-1}$, respectively. The transition dipoles of $\nu_{s}$(CF$_{2}$) at 1182 cm$^{-1}$ are in the same direction in $\alpha$ and $\delta$-phase unit cell and hence results no change in intensity at 1182 cm$^{-1}$ and it remains constant. While the transitional dipole moments of $\nu_{as}$(CF$_{2}$) at 1209 cm$^{-1}$, are in the opposite direction in unit cell of $\delta$-phase as compared to unit cell $\alpha$-phase. Thus, as a result 1209 cm$^{-1}$ absorption band does not appear in IR spectra of $\delta$-phase. \cite{doi:10.1021/acs.macromol.0c02567} This observation is consistent with FTIR spectra of PVDF nanoparticles processed via electrospray. Additionally, the FTIR spectra (Fig. S3 (marked as $\alpha$), supplementary material) also confirms the phase change, occurring from initial $\delta$-phase to the $\alpha$-phase due to the heat treatment at 170ºC. The scanning electron microscopy (SEM) of the $\delta$-PVDF nanoparticles is illustrated in Fig. 1(c), shows the formation of nanoparticles, ranges between $\sim$ 60 - 250 nm. The 2D and 3D AFM topography image (Fig. S4(a) and Fig. S4(b), supplementary material) displays the particles are spherical in shape with typical diameter around 90 nm.
\begin{figure}[h]
\includegraphics[scale=0.46]{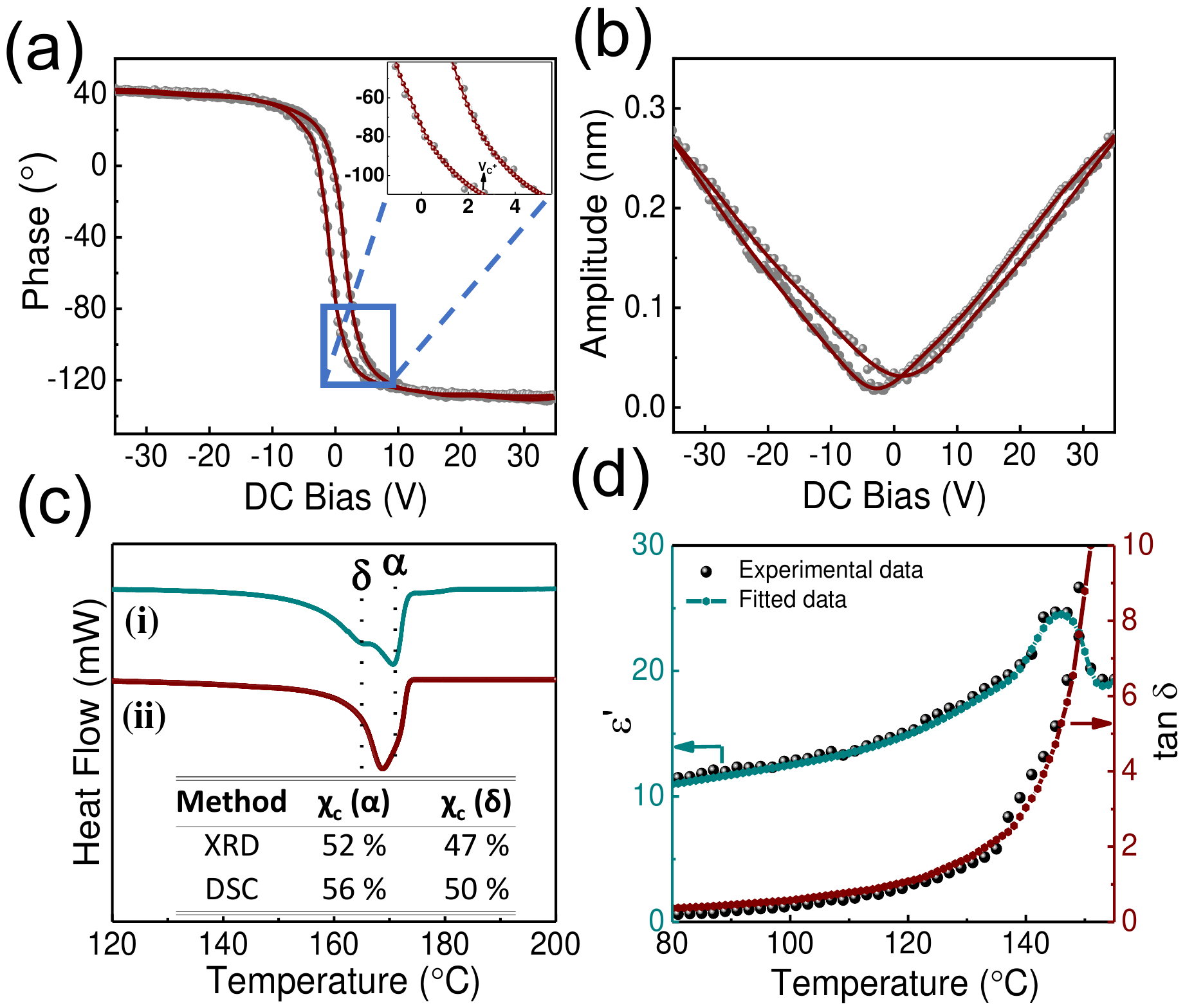}% Here is how to import EPS art
\caption{\label{fig:epsart} (a) Phase (b) Amplitude response of $\delta$-PVDF nanoparticles in PFM, after applying a DC bias of $\pm$ 35 V. Temperature dependent study of $\delta$-PVDF nanoparticles as (c) Heat flow in DSC at 1ºC/ min. followed by (i) first heating (ii) second heating cycle (inset shows the estimated crystallinity from DSC and XRD). (d) Dielectric constant ($\epsilon^{'}$) and loss tangent (tan $\delta$) of $\delta$-PVDF for variable temperature at 1 kHz frequency.}
\end{figure}
 
To study the piezoelectric and ferroelectric properties of $\delta$-PVDF nanoparticles, piezoresponse force microscopy (PFM) was performed. An AC modulating voltage was applied between the conductive probe and ITO coated substrate to induce the mechanical oscillation normal to the surface in contact mode. The lock-in drive frequency of 12 kHz with drive amplitude of 8 V was employed to get the proper amplitude (A) and phase response ($\phi$) of $\delta$-PVDF nanoparticles. The DC bias voltage of +35 V to -35 V was applied to obtain the corresponding phase and amplitude response of $\delta$-PVDF nanoparticles. The hysteresis plot (phase response) in Fig. 2(a) shows that the dipoles of the $\delta$-PVDF nanoparticles are reorienting itself under the application of external bias voltage. As the positive bias voltage is applied, the dipoles inside the crystalline lamella of $\delta$-PVDF nanoparticles are being aligned in the opposite direction of applied electric field as negative phase response. The maximum reorientation takes place until the bias voltage reaches to +35 V, with maximum dipole switching with phase of -130º and therefore saturation takes place. Now, as the bias voltage starts decreasing and it crosses the coercive voltage (V$_{c}$), the dipoles start reorienting itself and keeps reorienting till it saturates to other end with bias voltage of -35 V at 180º phase difference from the earlier phase. The coercive voltage (V$_{c}$) was found to be $\pm$ 2.5 V. This final hysteresis plot verifies the ferroelectric nature of these $\delta$-PVDF nanoparticles. The butterfly loop (amplitude response) shown in Fig. 2(b), is the evidence of piezoelectric response. As the bias voltage crosses the coercive voltage, the crystal lattice expansion takes place and it attains the maximum deformation of 0.27 nm at +35 V bias. As the applied DC voltage decreases, then it starts contracting itself until it reaches the minimal amplitude position, and start further expanding after increase in the magnitude of negative bias voltage above V$_{c}$ and reaches to almost similar maximum deformation amplitude, before it start decreasing with decrease in the magnitude of negative bias voltage. In this case, nanoparticle attains minimal position lower than the previous amplitude. The butterfly loop confirms the piezoelectric response of ferroelectric $\delta$-PVDF nanoparticles. The magnitude of \textit{d$_{33}$} value was calculated from the slope of the butterfly loop. \cite{al1985equilibrium} The obtained \textit{d$_{33}$} value for  $\delta$-PVDF nanoparticles is $\sim$-11 pm/V. The negative sign in this piezoelectric coefficient is associated with the direction of applied electric field with respect to the direction of polarization arising in the material. \cite{burkard1974reversible}

 The differential scanning calorimetry (DSC) was performed at the ramping rate of 1ºC /min under N$_{2}$ environment (Fig. 2(c)). The first heating cycle (Fig. 2(c-i)) from 120ºC - 200ºC, indicates the melting of different phases of PVDF. In this cycle, two peaks arise due to presence of two different phases of PVDF \cite{garcia2017ferroelectric}, which could be predicated as $\delta$ and $\alpha$-phase melting at 165ºC and 170ºC, respectively. In the cooling cycle, the sample crystallization takes place. Finally, the 2$^{nd}$ heating cycle ((Fig. 2(c-ii)) indicates only a single melting peak at 169ºC, that belongs to a single phase of PVDF. It can be interpreted as presence of $\alpha$-phase only, since this observation is consistent with the XRD (Fig. 1(b)) and FTIR (Fig. S3, supplementary material) results obtained from $\delta$-phase to $\alpha$-phase transformation when temperature is employed. The degree of crystallinity ($\chi_{c}$) was calculated (Discussion S5, supplementary material) from DSC thermogram and it is consistent with estimated crystallanity from XRD (Fig. 2(c) inset). The subtle difference in crystallinity value arises as compared to XRD, due to the melting enthalpy provided to polymer chain during DSC process, which results the reorientation of molecular chain and changes the overall crystalline region in polymer chain. The dielectric constant and loss tangent measurements of $\delta$-PVDF nanoparticles were performed at 1 kHz frequency with a temperature range of 80 to 155ºC has shown in Fig. 2(d). At room temperature (RT) and 1 kHz, the dielectric constant ($\epsilon^{'}$) for $\delta$-PVDF nanoparticle was measured $\sim$7.5, with the loss tangent (tan$\delta$) $\sim$0.25. For an increasing temperature range, the dielectric constant and loss tangent decreases with increase in frequency (Fig. S5, supplementary material). On the other hand, at a constant frequency (1 kHz), the dielectric constant and loss tangent increases gradually as the temperature increases. At lower temperature, these dipole molecules do not possess the flexibility to reorient themselves in the polymer chain. However, as the temperature increases, the possibility of dipoles reorientation results the increases in dielectric constant. In this scenario, the dielectric response is governed by the anomalous behavior of the glass transition of the polymer, but in the further increase in temperature results into the intense thermal vibrations that leads to suppress the degree of reorientation, resulting the reduction in dielectric constant of the polymer. \cite{el2014dielectric} This particular behavior in this temperature region could be explained as ferroelectric to paraelectric like phase transition in $\delta$-PVDF.
 
 The piezoelectric $\delta$-PVDF nanoparticles were obtained at significantly lower electric field (0.1 MV/m) in comparison to earlier reported work. The reason for the formation of $\delta$-phase nanoparticles, is majorly governed by Taylor cone \cite{vaseashta2007controlled} formation in electrospray process, in the presence of applied external electric field.\cite{ryan2012influence} However, the reason of low electric field driven $\delta$-phase formation through 180º chain rotation, can be explained by kink-propagation model. \cite{{PhysRevB.21.3700},{doi:10.1063/1.327422}} It also demonstrates that at lower electric fields, this phase transformation is significantly dependent on the temperature and the kink propagation velocity. To explain the kink rotation mechanism of TGTG´ chain unit by 180º, the motion of the chain has been defined by the following model. \\
The rotational kinetic energy of the molecular chain is defined as
\begin{equation}
T = \dfrac{1}{2}\ I\ \sum_{i} \biggl( \frac{\partial{\theta_i}}{\partial t} \biggr)^2
\end{equation}
where, I is moment of inertia of monomer unit about center of mass axis of the chain; $\theta_i$ is the rotation angle of $i^{th}$ monomer unit from the a-axis of the crystal, for a given time t.\\ 
The potential energy of the molecular chain is defined as
\begin{equation}
U = \sum_{i}
 \ [\ A_{1} (1- \cos {\theta_i})+A_{2} (1- \cos {2\theta_i})+ \frac{1}{2}\kappa ({\theta_i}- {\theta_{i+1}})^2\ ] 
\end{equation}
where, $A_1$ and $A_2$ are constants associated with dipole moment of TGTG´ unit and $\kappa$ is torsional constant. The first two terms (in Eq. 2) denote the combined inter-chain interaction potential with external electric field and last term appears as potential energy due to the torsional rigidity of the molecular chain.\\
The Hamiltonian for this modelled system can be written as
\begin{multline}
H= \dfrac{1}{2}\ I\ \sum_{i} \biggl( \frac{\partial{\theta_i}}{\partial t} \biggr)^2 + 
\sum_{i} \ [\ A_{1} (1- \cos {\theta_i}) + A_{2} (1- \cos {2\theta_i}) \\ + \frac{1}{2}\kappa ({\theta_i}- {\theta_{i+1}})^2\ ] 
\end{multline} 
Taking the Brownian motion of adjacent chains and anharmonic phonon forces into consideration for energy transfer,  the modelled system can be approximated to Langevin equation.\cite{doi:10.1063/1.327422} The approximation to minimize the thermal contribution force term into the continuum limit, results a generalized double Sine - Gordon equation: 
 \begin{equation}
I \ \frac{\partial^2{\theta}}{\partial t^2} = - A_{1} \ \sin {\theta} - 2A_{2} \ \sin {2\theta} + \kappa c^{2} \ \frac{\partial^2{\theta}}{\partial x^2} - \lambda I \ \frac{\partial {\theta}}{\partial t}
\end{equation}
where, $\lambda$ is the damping constant and c is the periodicity along chain axis. The possible solution to the above equation can be written as
 \begin{equation}
\theta (x,t) = 2 \ \arctan \ \exp \biggl(\frac{2(x + vt)}{d}\biggr)
\end{equation}
where, d is the width of the kink. The above equation represents the motion of the soliton wave of electric polarization. For $\theta$ = 0 to $\theta$ = $\pi$ at any time t, $\theta$(x,t) gives the probable propagation direction of kink that could travel from unpoled state to the poled state under the influence of applied electric field with kink velocity \textit{v}. Eventually, resulting as a chain segment rotation by 180 degrees. i.e. the possible favorable pathway to achieve the 
$\delta$-phase in PVDF chain by rotating its alternate second chain around c-axis.
\begin{figure}
\includegraphics[scale=0.25]{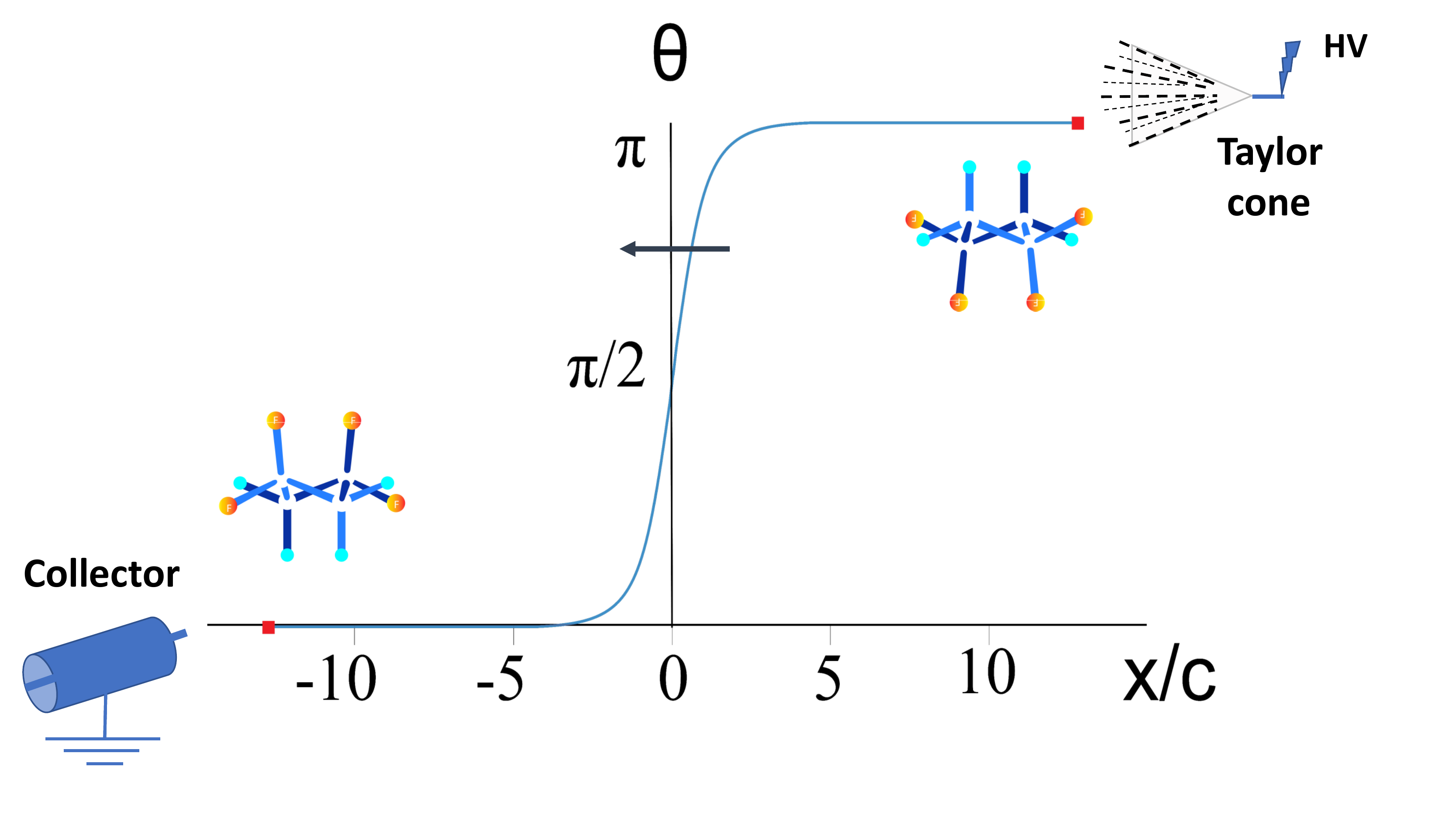}% Here is how to import EPS art
\caption{\label{fig:epsart} Kink propagation model for $\delta$-phase conversion in the electrospray system demonstrated by double Sine-Gordon solution. The ordinate shows the x/c, i.e. the distance (x) traveled by kink along the c-axis of unit cell and   abscissa shows the kink rotation angle at any time given t, in presence of electric field generated through electrospray system (Fig. S2). The beginning of the kink (right red dot) shows initial orientation of the polymer chain at taylor cone and at the final stage of the kink (left red dot) shows the 180º rotated PVDF polymer chain.}
\end{figure}

Now, following the continuum approximation, the propagation speed of kink along c-axis is given by
 \begin{equation}
v = v_o [ 1 + ({E_o}/{E})^2 ]^{-{{1}/{2}}}
\end{equation}
where, $v_o$ is the max kink velocity at larger electric field $E_o$.\cite{{doi:10.1063/1.327422}} As discussed by DveyAharon \textit{et al.} that at larger electric fields, the kink velocity doesn't depend on temperature. But at lower electric fields, the temperature plays a significant role to achieve the sufficient kink velocity for rotation. In our case, $E_o$ has been approximated to the limit to achieve a stable Taylor cone formation \cite{kim2007controlled}, i.e. $\sim$0.1 MV/m, for $\delta$-phase in electrospray system. We have predicted an approximated pathway, for $\delta$-phase conversion in electrospray process (Fig. 3) by simulating the kink propagation model for a lamella size of $\sim$12 nm.\cite{martin2017solid} Also, it is notable that, in  Taylor cone formation, the polymer is initially present in gaseous state, associated with molecular chains having higher entropy and free energy. Eventually, leading the kink propagation at very much faster rate as compare to thin films, in presence of electric field of same strength. So, this might be further reason for even lesser field requirement to achieve the $\delta$-phase in electrospray process. This theoretical model significantly predicts the phase conversion into $\delta$-phase at lesser electric field that observed experimentally.

Furthermore, $\delta$-PVDF nanoparticle based nanogenerator (NG) was prepared (Discussion S6, supplementary material), to show the application as an excellent piezoelectric energy harvester. The mechanical energy harvesting mechanism by the $\delta$-PVDF nanoparticles made device was explored by the finite element method (FEM) based theoretical simulation. It has observed that the NG generates higher stress (Fig. 4(a)) and thus higher piezo-potential (Fig. S6, supplementary material), in comparison that of planar PVDF film based device, due to stress confinement effect under application of 100 kPa pressure. The energy harvesting performance of NG was experimentally recorded in terms of open circuit output voltage (V$_{oc}$) and short-circuit current (I$_{sc}$) by repetitive finger imparting motion with the applied force amplitude of $\sim$4.5 N. The peak to peak response generated (Fig. 4(b)) as V$_{oc}$ $\sim$2.8 V and I$_{sc}$ $\sim$1.4 $\mu$A . The acoustic response of the NG was measured (with 40 W RMS) at $\sim$80 dB of sound pressure level (SPL) with tunable sound frequencies (40-140 Hz) (Fig. S7, supplementary material) indicating acoustic sensitivity of $\sim$2.75 V/Pa. This fact indicates that $\delta$-PVDF nanoparticles are in-situ poled during the electrospray process, thus further electrical poling step is not required like typical piezoelectric based sensor, actuators and energy harvesters. \\
\begin{figure}[h]
\includegraphics[scale=0.42]{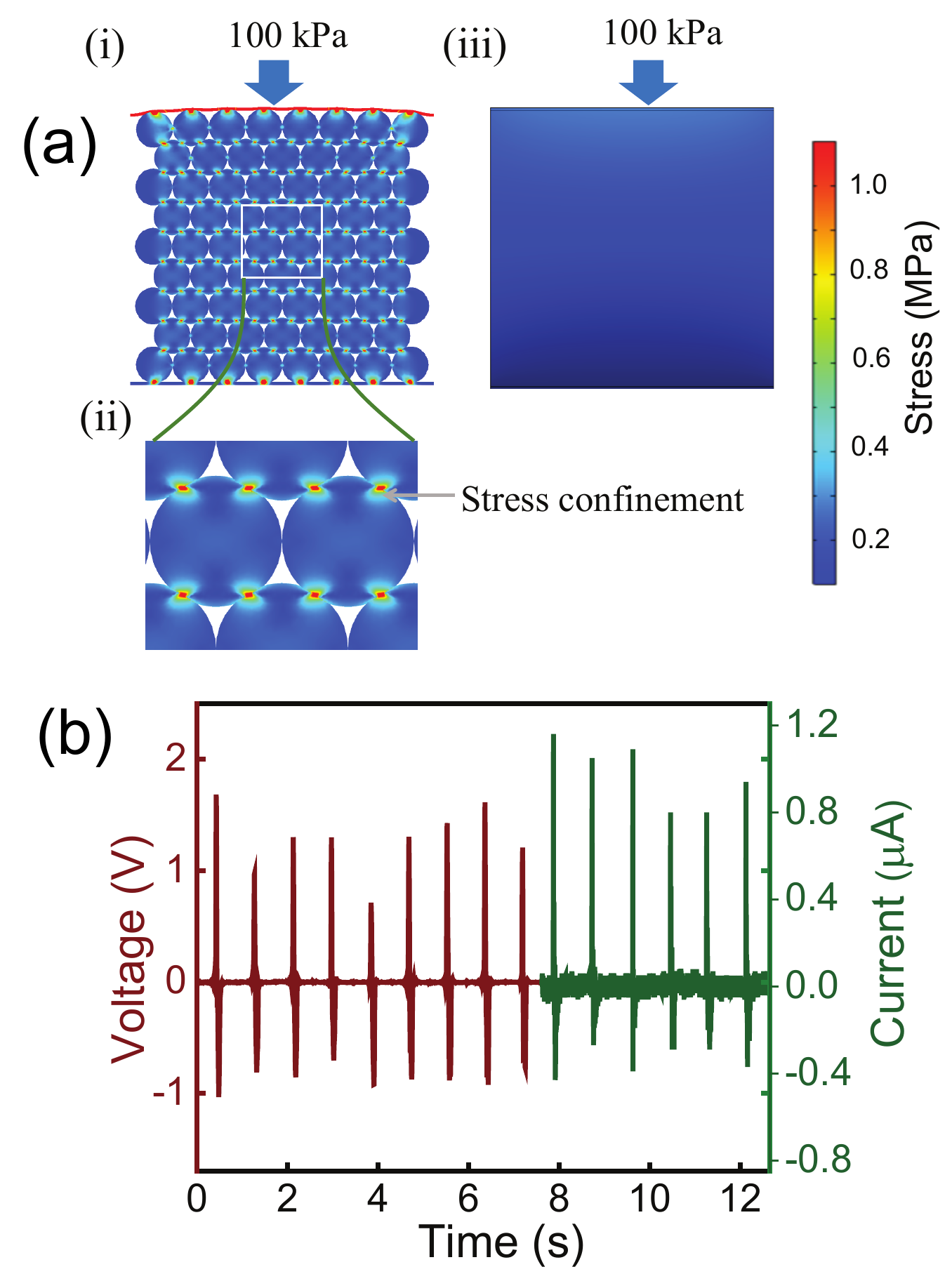}
\caption{\label{fig:epsart} FEM based theoretical simulation of piezoelectric response in NG made with $\delta$-PVDF nanoparticles. (a) Stress distribution of (i) nanoparticle based $\delta$-PVDF NG under 100 kPa pressure. (ii) stress confinement between the interfacial area of nanoparticles and (iii) stress-distribution of the planar $\delta$-Phase NG showing no variation of stress throughout the film. Piezoelectric response of fabricated NG as (b) Open circuit output voltage (V$_{oc}$) and short-circuit current (I$_{sc}$).}
\end{figure} 
In summary, we conclude that $\delta$-phase comprising PVDF nanoparticles were successfully fabricated through electrospray process under 0.1 MV/m of electric field. The ferro- and piezo- electric response of the $\delta$-PVDF nanoparticle is evident from saturated phase response and from butterfly response in PFM, respectively. The piezoelectric coefficient \textit{(d$_{33}$)} was obtained $\sim$-11 pm/V.  The phase transformation study indicates that the $\delta$-phase has relatively lower melting temperature of 165ºC than the $\alpha$-phase PVDF. Ferroelectric to paraelectric phase transition is also observed via DSC study. The $\delta$-phase transformation in electrospray process can be predicated by kink propagation model and can justify the low electric field requirement. The fabricated nanogenerator has shown a very prompt response with output voltage of $\sim$2.8 V and $\sim$1.4 $\mu$A current and can be useful in prospective applications for self-powered wearable electronics and sensors.\\

See supplementary material for experiment procedure, materials and methods. 

\begin{acknowledgments}
This work was financially supported by a grant from DST (DST/TMD/MES/2017/91), Government of India. V. G. is thankful to DST for awarding INSPIRE fellowship (IF180159).
\end{acknowledgments}

\nocite{*} 
\bibliography{MS}
\end{document}

\begin{center}
{\LARGE Supplementary Materials}
\end{center}

\title[]{{\hspace{10.0mm} $\delta$-PVDF Based Flexible Nanogenerator}}

\author{Varun Gupta}
\affiliation{Quantum Materials and Devices Unit, Institute of Nano Science and Technology, Knowledge City, Sector 81, Mohali 140306, India}
\author{Anand Babu}
\affiliation{Quantum Materials and Devices Unit, Institute of Nano Science and Technology, Knowledge City, Sector 81, Mohali 140306, India}
\author{Sujoy Kumar Ghosh}
\affiliation{Department of Physics, Jadavpur University, Kolkata 700032, India}
\affiliation{Present address: NEST, Istituto Nanoscienze-CNR and Scuola Normale Superiore, Piazza San Silvestro 12, I-56127 Pisa, Italy}
\author{Zinnia Mallick}
\affiliation{Quantum Materials and Devices Unit, Institute of Nano Science and Technology, Knowledge City, Sector 81, Mohali 140306, India}
\author{Hari Krishna Mishra}
\affiliation{Quantum Materials and Devices Unit, Institute of Nano Science and Technology, Knowledge City, Sector 81, Mohali 140306, India}
\author{Dipankar Mandal}
\email{dmandal@inst.ac.in}
\affiliation{Quantum Materials and Devices Unit, Institute of Nano Science and Technology, Knowledge City, Sector 81, Mohali 140306, India}

\maketitle 

\hspace{-5.0mm}\textbf{Associate discussion S1: $\delta$-phase in PVDF} \vspace{1.5mm}  \\ 
There have been reported very few different techniques to achieve the $\delta$-phase in PVDF, as mentioned in Fig. S1. Most of them required the higher electric field for $\delta$-phase formation, likely more than 100 MV/m. The electric field strength required in this work is 10$^{3}$ times lesser than the earlier reported work (Table SI). To our best of knowledge, this is very first report of formation of $\delta$-phase comprising piezoelectric PVDF nanoparticles under such lower electric field of 0.1 MV/m at room temperature in ambient conditions.\\

\setcounter{table}{0}
\renewcommand{\thetable}{S\Roman{table}}%
\setcounter{figure}{0}
\renewcommand{\thefigure}{S\arabic{figure}}%

\begin{figure}[h]
\includegraphics[scale=0.60]{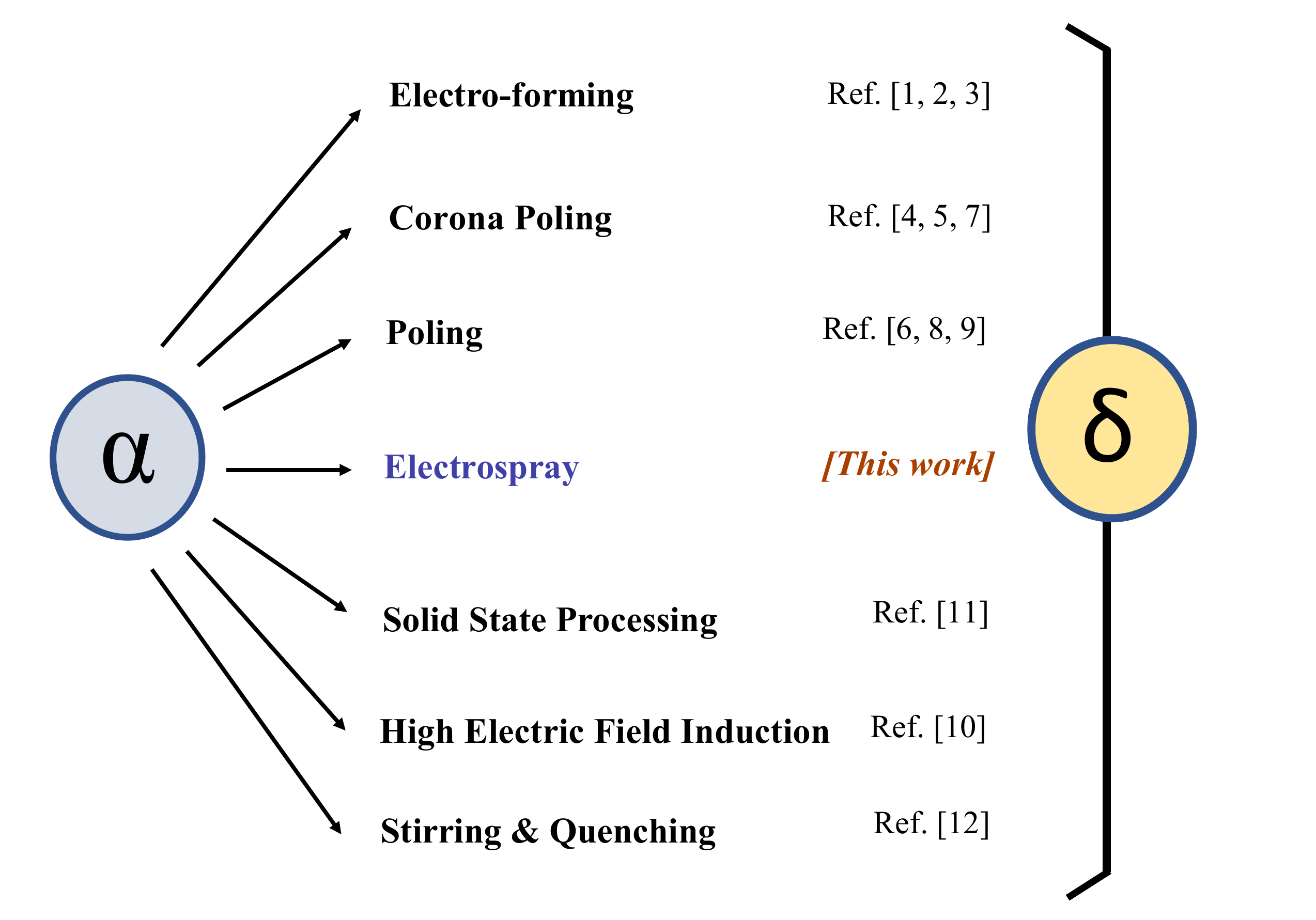}% Here is how to import EPS art
\caption{\label{figure:epsart} Different reported techniques to obtain the $\delta$-phase in PVDF.} 
\end{figure}
\newpage
\begin{figure}
\includegraphics[scale=0.65]{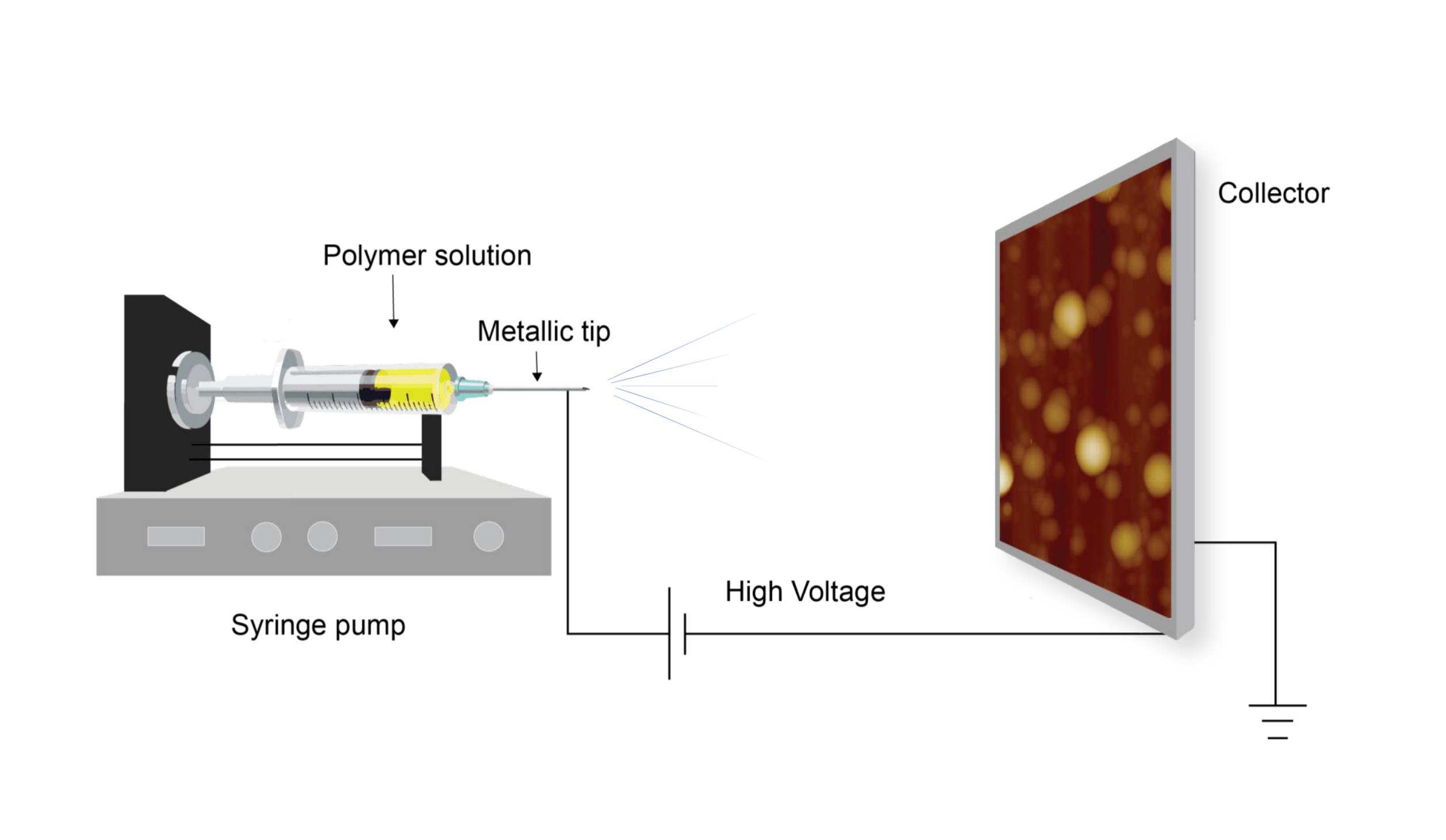}% Here is how to import EPS art
\caption{\label{fig:epsart} Schematic of electrospray setup, where a high voltage is applied between metallic tip and collector.}
\end{figure}
\newpage
\begin{table*}
\caption{\label{tab:table3} Electric field strength and other details for achieving $\delta$-phase in PVDF.}
\begin{ruledtabular}
\begin{tabular}{cccccc}

   {\makecell {Sample type \\(Technique used)}}  &  {\makecell { Temperature \\ (ºC) }}  & {\makecell {Electric Field \\ strength (MV/m)}} & {\makecell {References \\}}\\
\hline

 {\makecell *[{{p{6.2cm}}}]{Biaxially oriented thick film, t $\sim$ 12 $\mu$m (Electro-forming).}} & 25  & 170 & \cite{davis1978electric}\\
  {\makecell *[{{p{6.2cm}}}]{Thin film, t $\sim$ 400 nm \\ (Electro-forming).}} & 25  & 250 & \cite{{li2013revisiting},{li2013low}}\\
  {\makecell *[{{p{6.2cm}}}]{Drawn film, t $\sim$ -\footnotemark[1] $\mu$m\\ (Corona poling).}} & 25  & 200 & \cite{bachmann1980crystal}\\
  {\makecell *[{{p{6.2cm}}}]{Thick film, t $\sim$ 20 $\mu$m \\  (Constant current corona charging).}} & -\footnotemark[1] & 220 & \cite{costa1993electric}\\
  {\makecell *[{{p{6.2cm}}}]{Stretched film, t $\sim$ -\footnotemark[1] $\mu$m \\ (DC Poling).}} & 24  & 150  & \cite{naegele1978formation}\\
 {\makecell *[{{p{6.2cm}}}]{Thin film, t $\sim$ 400 nm\\ (Corona poling).}} & 25 & 500 & \cite{gan2016ferroelectric}\\
  {\makecell *[{{p{6.2cm}}}]{Drawn thick film, t $\sim$ 100 $\mu$m \\ (DC poling).}} & 80 & 100  & \cite{servet1979polymorphism}\\
 {\makecell *[{{p{6.2cm}}}]{Drawn thick film, t $\sim$ 100 $\mu$m \\ (Poling)}} & 130 & 100  & \cite{davies1979evidence} \\
 {\makecell *[{{p{6.2cm}}}]{Biaxially drawn thick film, t $\sim$ 10 $\mu$m \\ (High electric field induction).}} & -\footnotemark[1] & 300 & \cite{doi:10.1021/acs.macromol.0c02567} \\
 {\makecell *[{{p{6.2cm}}}]{\textbf{Nanoparticles}, (Diameter: 60 - 250 nm) \\ \textbf{(Electrospray)}.}} & \textbf{\textit{25}} & \textbf{ 0.1 } & \textbf{\textit{This work}}\\
\footnotetext[1]{Data not available}

\end{tabular}
\end{ruledtabular}
\end{table*}

\hspace{-5.0mm}\textbf{Associate discussion S2: Experimental section} \vspace{1.5mm}  \\ 
PVDF pallets (Aldrich, USA, Mw: 180,000 by GPC) were dissolved in N, N- dimethylformamide (DMF, Merck, India) solvent and stirred at 60ºC for 24 hours to obtain the final 12 wt\% PVDF solution for electrospray. The prepared PVDF solution was further loaded into a 10 ml syringe and mounted to electrospray system. The electrospray process was performed at the optimum voltage of 12 kV to achieve the $\delta$-PVDF nanoparticles. The distance between the metallic tip to collector was kept constant as 12 cm, followed by sample flow rate of 0.4 mL/hr. The final sample was collected on an aluminium foil wrapped over the plate collector. During the whole experimental procedure, the relative humidity present in chamber was  $\sim$50\%, at room temperature (25°C).\\\\
\newpage 
.\\
\textbf{Associate discussion S3: Characterization} \vspace{1.5mm}  \\ 
High-resolution transmission electron microscopy (HRTEM) obtained at 200 kV, selected area electron diffraction (SAED) and elemental mapping analysis were carried out by JEOL, JEM-2100 electron microscope. The morphology was performed using scanning electron microscopy (SEM JEOL JSMIT 300). X-ray diffraction (XRD) study was performed on a Bruker D8 advances diffractometer using Cu-$K_{\alpha}$ ($\lambda$ $\sim$1.5418 {\AA}) radiation in the 2$\theta$ range from 10° to 35° with an acceleration voltage of 40 kV. Fourier transform infrared spectroscopy (FTIR) was measured in ATR mode with spectral resolution of 4 cm$^{-1}$, from 500 - 4000 cm$^{-1}$ wavenumber by iS5 Nicolet Thermofisher scientific instruments. The surface topography was measured by Atomic force microscopy (AFM) Bruker Multimode-8 system using etched crystal silicon probe (spring constant $\sim$40 N/m) in non-contact mode, with probe oscillation frequency of 372 kHz. Piezoresponse force microscopy (PFM) was performed using the conductive Pt-Ir coated probe with spring constant $\sim$3 N/m in contact mode with sample deposited on conducting ITO coated glass substrate. Differential scanning calorimetry (DSC) was carried out by a Perkin Elmer STA-8000 instrument under N$_{2}$ environment. Dielectric measurements were performed on Keysight Impedance Analyzer E4990A system, where two parallel plate capacitor geometry is considered. \\
 
\hspace{-5.0mm}\textbf{Associate discussion S4: Crystallite size and degree of crystallinity ($\chi_{c}$) } \vspace{1.5mm}  \\ 
The crystallite size in the $\delta$-phase nanoparticle was calculated by Debye Scherrer formula 
 
 \begin{equation}
L =  \frac{k \lambda}{B \cos \theta} 
\end{equation}
 \\
where, \textit{L} is crystallite size, \textit{$\lambda$} is X-ray wavelength ($\sim$1.5418 {\AA}) of Cu-$K_{\alpha}$, $\theta$ is Bragg's angle, \textit{B} is line broadening (FWHM) and \textit{k} is shape factor ($\sim$0.89). The crystallite size of $\delta$-phase and $\alpha$-phase was found to be $\sim$6 nm and $\sim$13 nm, respectively.
Further, the percentage crystallinity of $\delta$-phase and $\alpha$-phase from XRD, was determined by the following equation  

 \begin{equation}
  \chi_{c} (\%) =  \frac{\Sigma  A_{cr}}{\Sigma A_{cr} + \Sigma A_{am}} \times  100 \%
\end{equation}
 \\
where, $\Sigma  A_{cr}$ is the total area of crystalline phase and $\Sigma  A_{am}$ is the total area of amorphous phase in deconvoluted XRD peaks. The crystallinity of $\delta$-phase and $\alpha$-phase was found to be $\sim$ 47\% and $\sim$ 52\% respectively.
\\\\

\hspace{-5.0mm}\textbf{Associate discussion S5: Degree of crystallinity ($\chi_{c}$) estimated from DSC thermogram} \vspace{1.5mm}  \\
The Degree of crystallinity from DSC thermogram was calculated from the following formula 
 \begin{equation}
\chi_{c} (\%) =  \frac{\Delta H_{m}}{\Delta H_{\infty}} \times  100 \%
\end{equation}
where, $\Delta H_{\infty}$ is standard enthalpy of fusion of fully crystalline sample, i.e. 104.6 J/g. $\Delta H_{m}$ is enthalpy of fusion in DSC thermogram. The crystallinity in $\delta$-PVDF nanoparticles estimated from DSC is $\sim$50\% and $\sim$56\% for $\alpha$-phase comprising PVDF.
 \\\\

\hspace{-5.0mm}\textbf{Associate discussion S6: Device fabrication and performance test} \vspace{1.5mm}  \\ 
Nanogenerator (NG) was made with $\delta$-PVDF nanoparticles layers of thickness $\sim$50 $\mu$m with an active area of 7.1 cm × 4.4 cm. Then top and bottom electrodes were prepared by using the conducting fabric with an active contact with the electrosprayed layer. Then both the electrodes were connected with copper wire to make the proper connection for electrical measurements (with the final device thickness of $\sim$200 $\mu$m). Further, the NG was coated with polydimethylsiloxane (PDMS) to protect from external damage and heat. The open circuit voltage and short circuit current were measured using Keysight DSO X1102G and Keysight source meter B2902A, respectively. The acoustic data was generated using i-ball Tarang Lion speaker (40 W), when NG is placed on the surface of the speaker and the output response were recorded in DSO.\\\\\\\\\\\\\\

\begin{figure}
\includegraphics[scale=0.65]{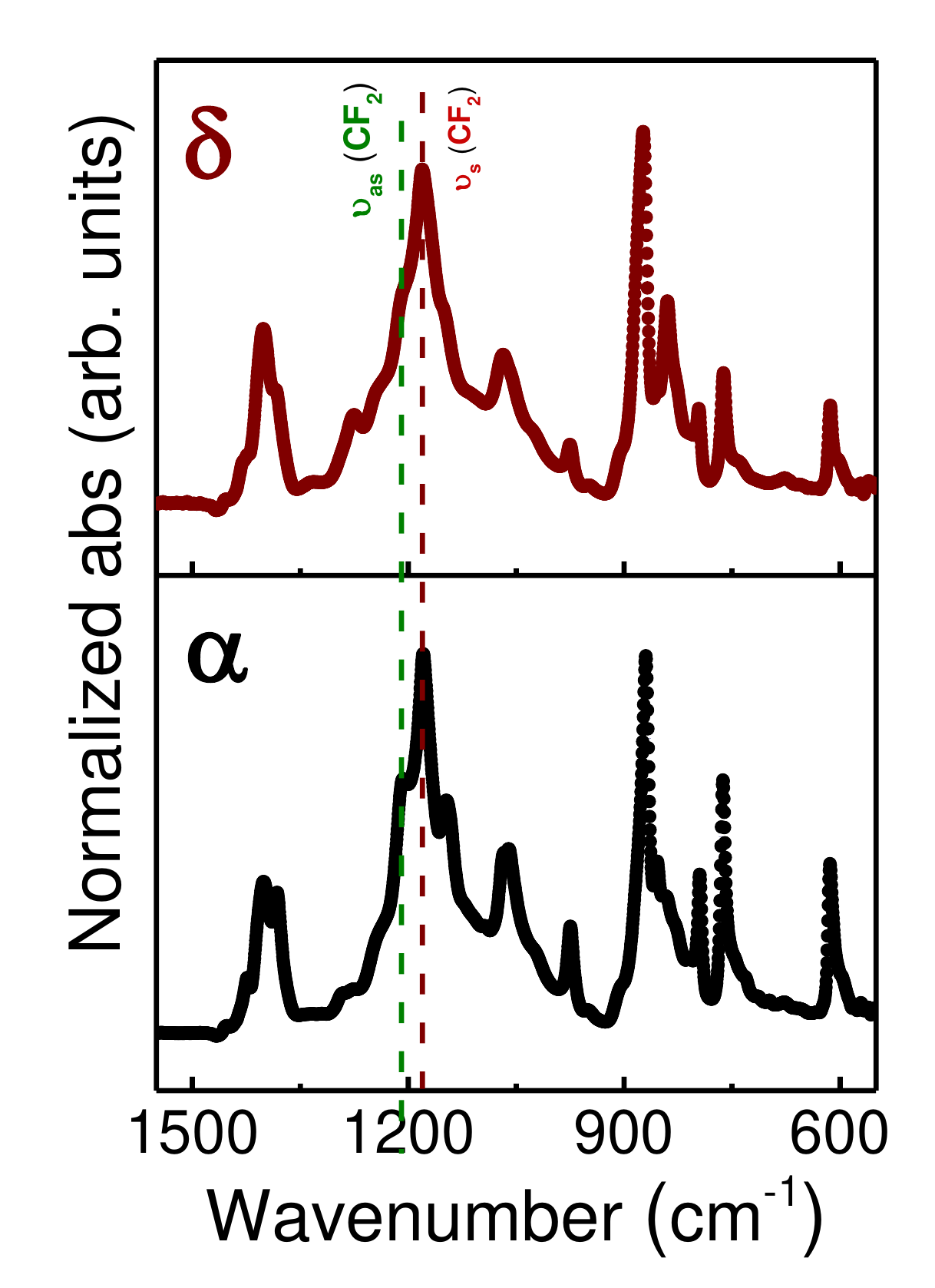}% Here is how to import EPS art
\caption{\label{fig:epsart} FT-IR spectra of PVDF nanoparticles with $\delta$ and $\alpha$-phase. The upper panel (marked as $\delta$) is for as prepared $\delta$-PVDF nanoparticles and the lower panel (marked as $\alpha$) is for $\alpha$-PVDF, when the phase transformation is attained by thermal treatment at 170ºC.}
\end{figure}
\begin{figure}
\includegraphics[scale=0.6]{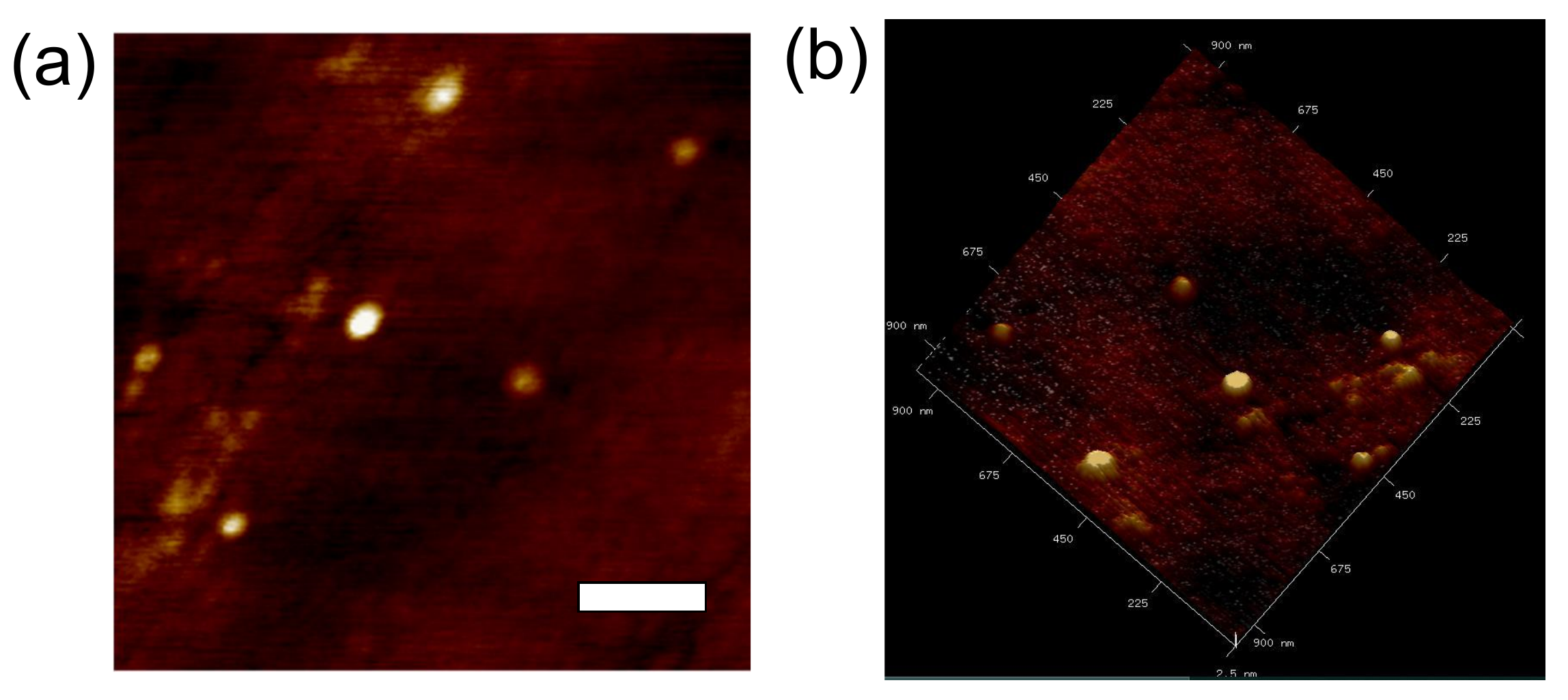}% Here is how to import EPS art
\caption{\label{fig:epsart}(a) AFM topography of $\delta$-PVDF nanoparticles with (a) 2D and (b) 3D view (scale bar 225 nm).}
\end{figure}
\begin{figure}
\includegraphics[scale=0.65]{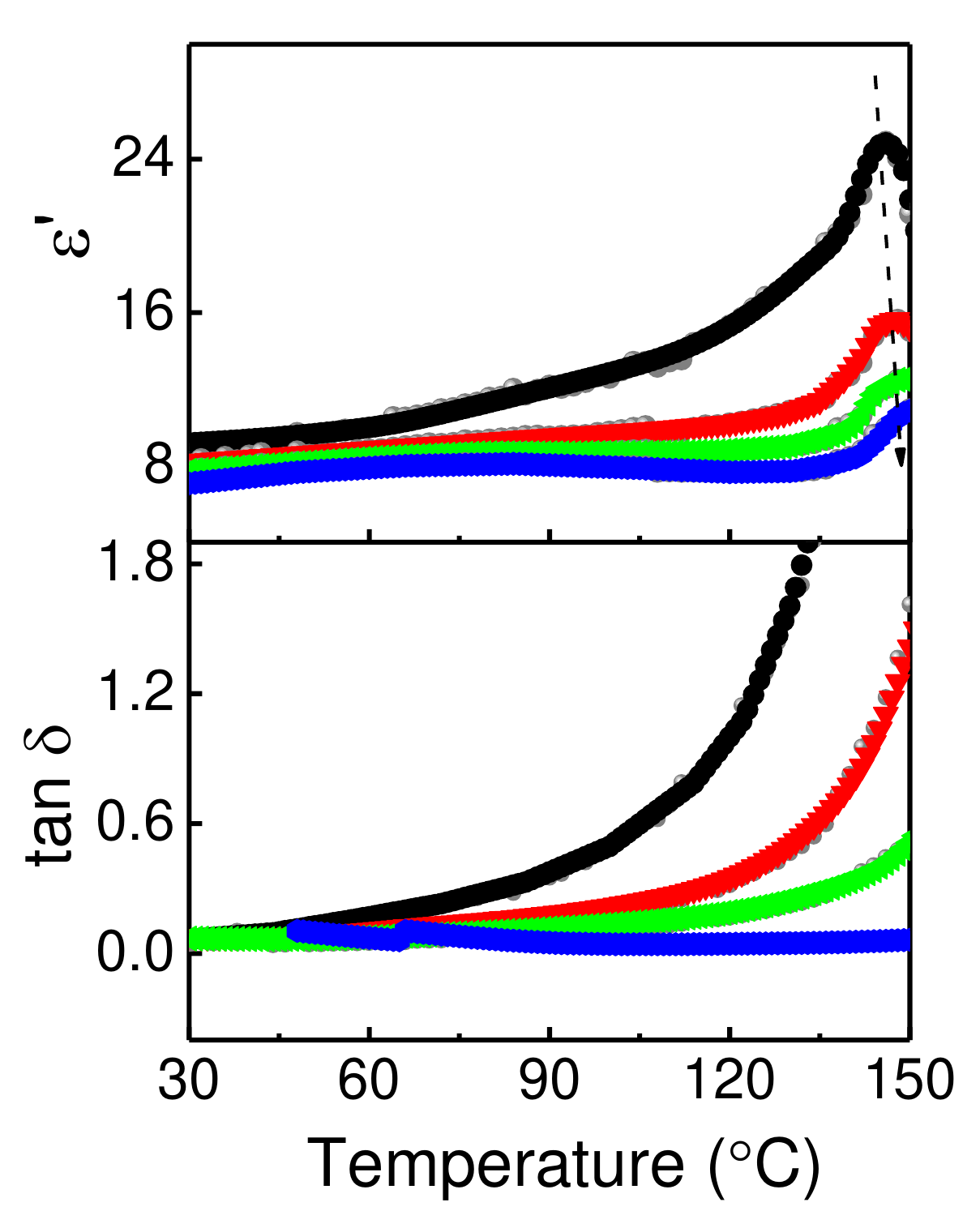}% Here is how to import EPS art
\caption{\label{fig:epsart} Dielectric constant ($\epsilon^{'}$) and loss tangent (tan$\delta$) of $\delta$-PVDF nanoparticles with variable temperature range of 30-155ºC at certain frequencies, viz., 1, 10 , 50 and 500 kHz.}
\end{figure}
\newpage
.\\\\\\\\\\\\\\\\\\
\begin{figure}
\includegraphics[scale=0.65]{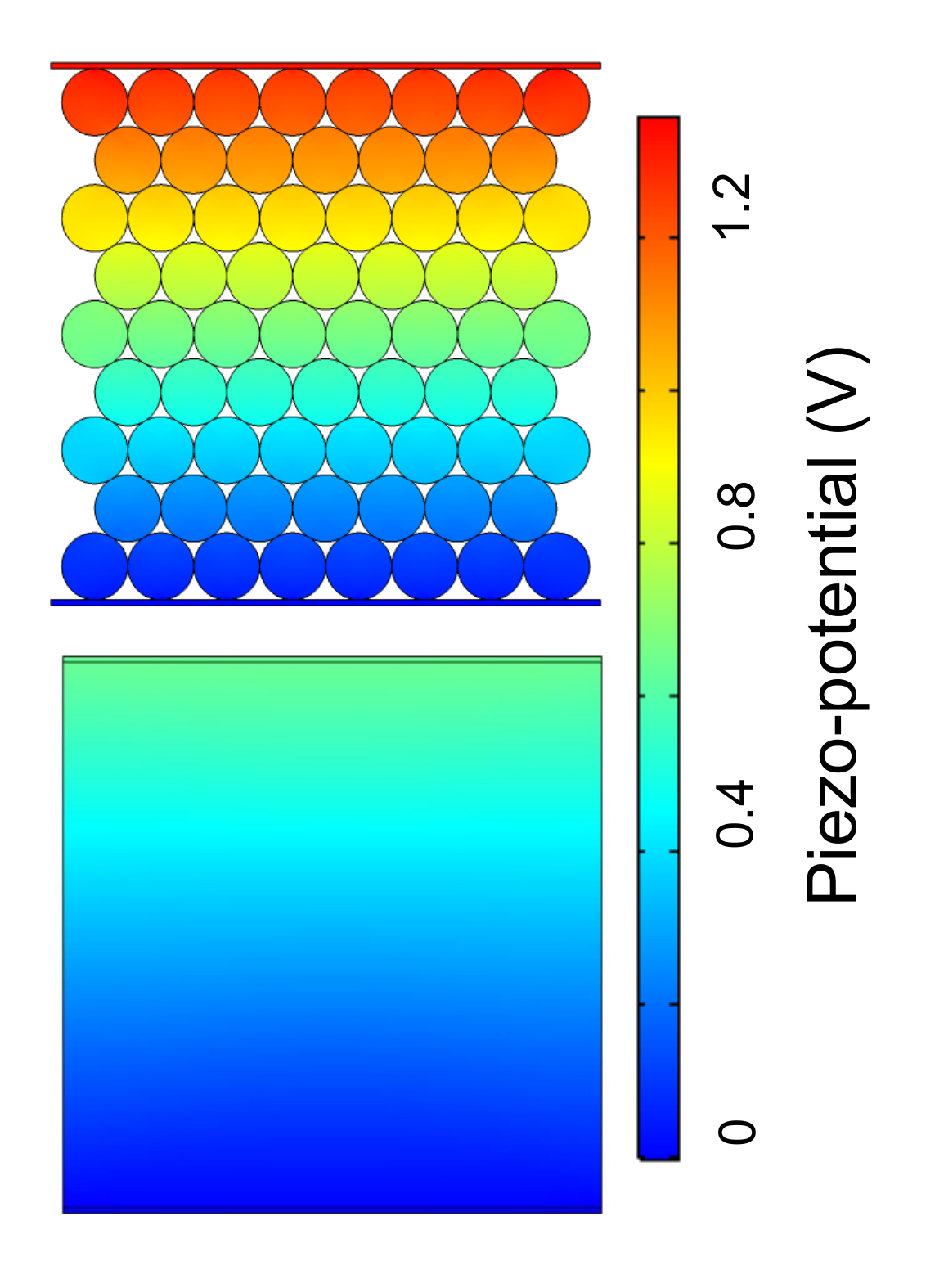}% Here is how to import EPS art
\caption{\label{fig:epsart} FEM based theoretical simulation for PVDF nanoparticles and planar film. It shows the relative generated piezo-potential under the application of 100 kPa pressure in both cases.}
\end{figure}
\begin{figure}
\includegraphics[scale=0.65]{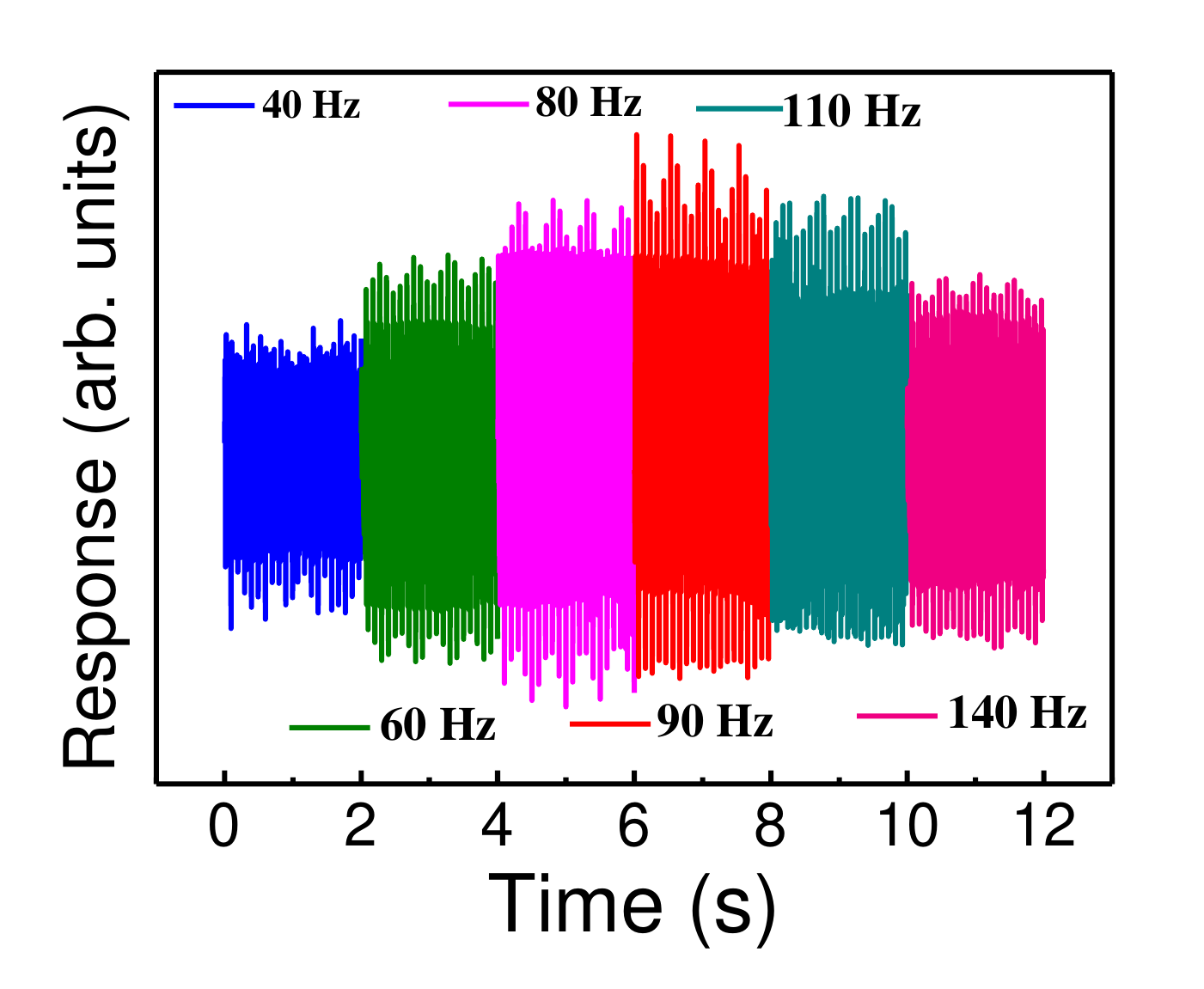}% Here is how to import EPS art
\caption{\label{fig:epsart} Acoustic response of NG at various frequencies with acoustic sensitivity of $\sim$2.75 V/Pa.}
\end{figure}
\newpage

\nocite{*}

\bibliography{Supple}